\documentclass[9pt, aip, preprint, amsmath, amssymb]{article}
\usepackage{siunitx}
\usepackage{graphicx}
\usepackage{authblk}
\begin{document}

\title{An integrated diamond Raman laser pumped in the near-visible}

\author[1]{Pawel Latawiec}
\author[1,2]{Vivek Venkataraman}
\author[1,3]{Amirhassan Shams-Ansari}
\author[4]{Matthew Markham}
\author[1,*]{Marko Lon\v{c}ar}

\affil[1]{John A. Paulson School of Engineering and Applied Sciences, Harvard University, Cambridge, MA, USA}
\affil[2]{Current affiliation: Department of Electrical Engineering, Indian Institute of Technology Delhi, New Delhi, India}
\affil{Department of Electrical Engineering and Computer Science, Howard University, Washington, DC, USA}
\affil[4]{Element Six Innovation, Fermi Avenue, Harwell Oxford, Didcot, Oxfordshire OX110QR, UK}

\affil[*]{Corresponding author: loncar@seas.harvard.edu}

%% To be edited by editor
% \dates{Compiled \today}

%% To be edited by editor
% \doi{\url{http://dx.doi.org/10.1364/XX.XX.XXXXXX}}

\maketitle

\begin{abstract}
Using a high-Q diamond microresonator (Q > 300,000) interfaced with high-power-handling directly-written doped-glass waveguides, we demonstrate a Raman laser in an integrated platform pumped in the near-visible. Both TM-to-TE and TE-to-TE lasing is observed, with a Raman lasing threshold as low as 20 mW and Stokes power of over 1 mW at 120 mW pump power. Stokes emission is tuned over a 150 nm (60 THz) bandwidth around 875 nm wavelength, corresponding to 17.5\% of the center frequency.

\end{abstract}

%\keywords{Raman Lasers; Microcavity Lasers, Nonlinear optics, Raman effect; Nonlinear optics, integrated optics; Integrated optics, Micro-optical devices; Nonlinear optics, Scattering, stimulated; Visible lasers.}

\section{Introduction}

% replace with intro paragraph on Raman instead of diamond?
As technologies for the manufacture of synthetic diamond crystals have matured, its use in and outside the laboratory has become more commonplace. Diamond is broadly transparent, passing light from the UV ($\SI{>220}{\nm}$) to $\SI{}{\THz}$ range, with the exception of losses present at $\sim2.6-\SI{6}{\um}$ due to multiphonon-induced absorption. Its superior thermal properties are well-known, with a thermal conductivity of $\SI[per-mode=symbol]{\sim1800}{\W\per\m\per\K}$ @ $\SI{300}{\K}$ and a low thermo-optic coefficient of $\sim10^{-5}\SI{}{\per\K}$\cite{Mildren}.

% A few words about diamond Raman lasers
Bulk Raman lasing systems have employed synthetic diamonds since their creation. Due to diamond's high-frequency optical phonons (39.99 THz) and large Raman gain, efficient \cite{Mildren2009a} and high power \cite{Williams2014} benchtop systems have been developed that span the UV \cite{Granados2011}, visible\cite{Mildren2008, Mildren2009a}, and infrared\cite{Sabella2011, Williams2014} wavelengths. The design freedom inherent in tabletop systems has spawned a number of innovations\cite{Kitzler2017, McKay2017}, although continuous-wave (CW) pumping and low thresholds remain difficult to reach, and benchtop components require careful alignment.

% A few words about integrated Raman lasers
The first experimental demonstrations of micro-scale Raman lasing occurred in silica microspheres and microtoroids\cite{Spillane2002, Kippenberg2004a}. Demonstrations in silicon\cite{Rong2005a} soon followed, showing cascaded Raman lasing\cite{Rong2008} and ultra-low thresholds in a photonic crystal geometry\cite{Takahashi2013}. Recent experimental demonstrations include aluminum nitride Raman lasers at telecom wavelengths\cite{Liu2017}. Following these platforms, both frequency combs\cite{Hausmann2014} and Raman lasers\cite{Latawiec2015} have been demonstrated in diamond microresonators.

% Difficulties in high-power handling, nonlinear photonics
Integrated photonic devices operating at visible wavelengths must overcome additional challenges relative to their telecommunications-band counterparts. Apart from relatively immature instrumentation, there is a dearth of qualified low-loss material platforms. This requirement is even more stringent in nonlinear photonic systems, where high visible pump powers and non-negligible linear and nonlinear absorption can lead to device failure. To enable pumping of the diamond microresonator system at sufficient powers above threshold, we develop a directly-writable doped-glass waveguide and integrate it with our diamond waveguides, creating an end-fire interface. Per these developments, we demonstrate the first integrated Raman laser pumped at near-visible, studying in particular polarization conversion between pump and Stokes and low-threshold lasing.

\begin{figure}[!htbp]
\centering
\includegraphics[width=8.4cm]{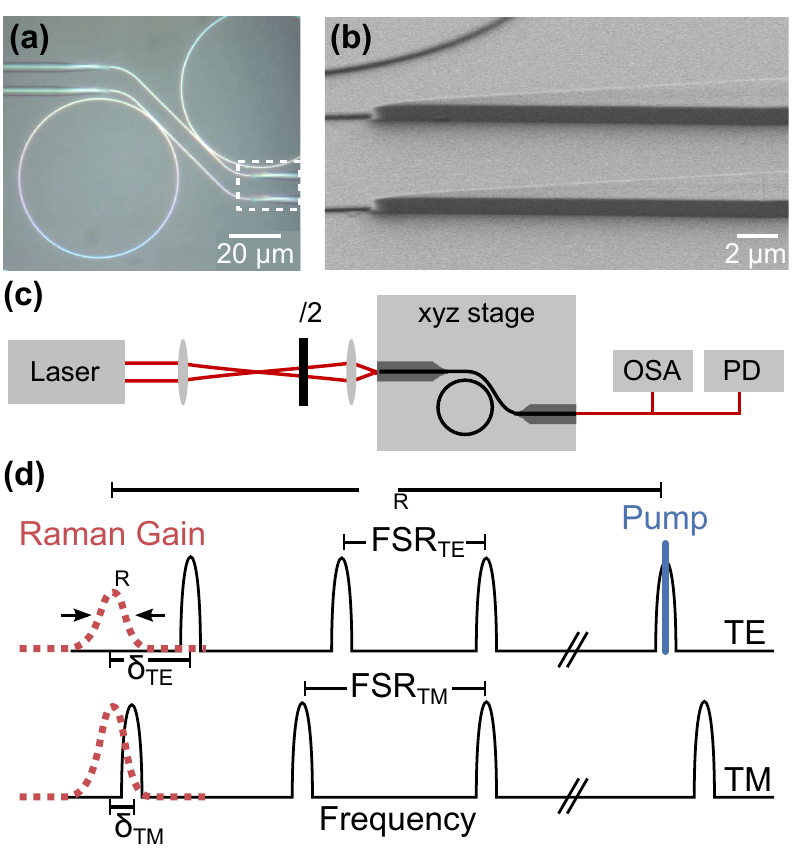}
\caption{Microresonator design and experimental concept. (a) Micrograph of diamond ring resonators, with coupling region highlighted. (b) Scanning electron micrograph of doped-glass to diamond coupling region. (c) Experimental setup. The light from a CW Ti:Sapphire laser is focused onto a waveguide facet after passing through a half-wave plate. A lensed fiber collects light and directs it to an optical spectrum analyzer and a photodetector. (d) Experimental concept. A CW pump is tuned to resonance, creating a Raman gain region at $\Omega_R \SI{\sim39.99}{\THz}$ Stokes shift and bandwidth $\Gamma_R \SI{\sim60}{\GHz}$ for both TE and TM polarizations. The strength of the Raman gain for these polarizations is dependent on the specific microresonator and diamond crystal orientation\cite{Mildren}. The difference from the Stokes Raman gain to the nearest resonance mode is termed the Stokes-mode mismatch, $\delta$.}
\label{fig:schematic}
\end{figure}

\section{Device Design and Fabrication}
% Diamond fabrication
Device fabrication broadly follows the same process reported previously\cite{Hausmann2012, Latawiec2015}. A 1x1 mm, $\SI{\sim30}{\um}$ thick electronic-grade single-crystal diamond with [001]-oriented surface (Element Six) was cleaned in a refluxing acid mixture. After rinsing in water, the diamond was then placed from methanol directly on a sapphire carrier wafer. This promotes a loose adhesion to the carrier. The diamond is etched on one side, then flipped, cleaned, and etched (Ar/Cl$_2$ cycled with O$_2$). The thinned diamond is then transferred to fused silica by first de-bonding from the carrier wafer with a drop of hydrofluoric acid, which is then gently washed away and diluted. Immediately before re-bonding to the silica substrate, the silica surface is activated with an O$_2$ plasma (300 mT, 100 W, 1 minute). Once the diamond is bonded, a monolayer of Al$_2$O$_3$ is deposited via atomic layer deposition (ALD), promoting adhesion of the FOx-16 electron-beam resist (spin-on glass, Dow Corning). The resist is written under multi-pass exposure (Elionix F-125) with waveguides aligned parallel to the diamond thickness gradient\cite{Latawiec2015}. The finished resonators (Fig. \ref{fig:schematic}(a)) are $\SI{60}{\um}$ in diameter and have cross-sectional dimensions of $\SI{300}{\nm}$ in width and $\SI{\sim 300}{\nm}$ in height. The diamond waveguides are adiabatically tapered down in width across a length of $\SI{200}{\um}$ to enable high coupling efficiency.

% Waveguide fabrication
The diamond sample is prepared for subsequent processing by depositing $\SI{\sim2}{\nm}$ of ALD oxide\cite{Hiller2010a}. This adheres the diamond to the substrate and prevents de-bonding. To prevent out-gassing when under high optical power, the sample is annealed at 460 C for 1 hour in O$_2$. This drives impurities out of the ALD film. Directly-written doped-glass waveguides to be used as spot-size converters (Fig. \ref{fig:schematic}(b)) are then defined with an electron beam exposure and developed (Appendix \ref{section:waveguidefab}).

Afterwards, a protective layer of photoresist is spun and a dicing saw is used to define a channel on the substrate's backside. The device is then cleaved, removed of resist, and cleaned. Finally, the sample is annealed at 460 C for 3 hours in an oxygen atmosphere to drive out residual impurities in the doped-glass waveguides and ALD film, as well as to etch away any final graphitized carbon and terminate the diamond surface in oxygen bonds\cite{Osswald2006}.

\section{Optical Measurements}

Optical measurements of the devices are performed in an end-fire testing setup, schematically shown in Fig. \ref{fig:schematic}(c). An aspheric singlet lens focuses a CW Ti:Sapphire laser (M2, Solstis) onto the facet of a waveguide. Free-space coupling has several advantages over a lensed optical fiber, as was used in previous experiments\cite{Latawiec2015}, by reducing insertion loss and maintaining stable polarization even at high powers.

After passing through the device, the light is collected on the other end with a lensed fiber (OZ Optics). A typical device showed losses of $\sim$5 dB/facet. An optical fiber splitter is then used to send the output to a photodetector for resonance measurements, or an optical spectrum analyzer (OSA, Yokogawa AQ6370) for Raman measurements.

\begin{figure}[htbp]
\centering
\includegraphics[width=8.4cm]{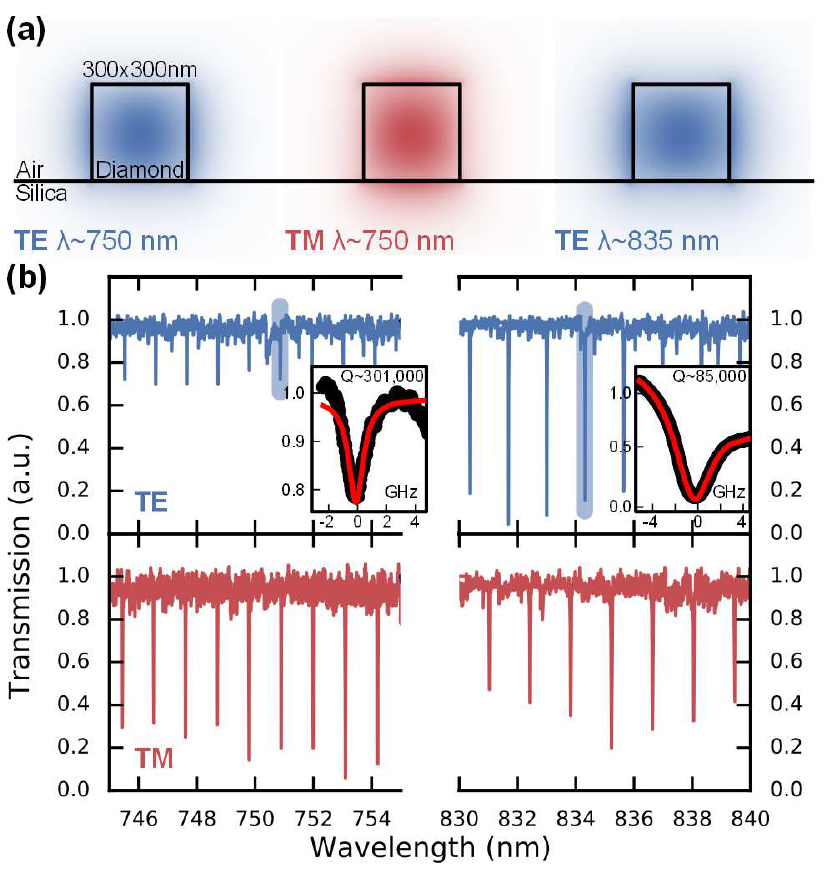}
\caption{Low-power characterization of Device 1. (a) Cross-sections of simulated mode intensities at TE pump (left, blue), TM pump (center, red) and TE Stokes (right, blue). (b) Transmission data for studied resonator at TE (top, blue) and TM (bottom, red) probe polarizations. Insets show zoomed-in highlighted regions of the TE spectra. The studied pump resonance shows a total quality factor $Q \sim$301,000 at $\SI{750.88}{\nm}$, corresponding to a Stokes resonance of $Q \sim$85,000 at $\SI{834.34}{\nm}$. The corresponding total $Q$ for the TM modes was 75,000 and 35,000, respectively.}
\label{fig:resonances}
\end{figure}

% Characterization details, Q factors
Two nominally equivalent devices are reported. The first device under test is characterized in Fig. \ref{fig:resonances}. Its waveguide cross-sections are shown in Fig. \ref{fig:resonances}(a). Both TE and TM spectra are shown in Fig. \ref{fig:resonances}(b), where the insets show the TE pump resonance used and its corresponding Stokes resonance. The TE pump (TM pump) and TE Stokes (TM Stokes) quality factors are 301,000 (75,000) and 85,000 (35,000), respectively, representing an order-of-magnitude improvement over the previous state-of-the-art for these wavelengths\cite{Hausmann2012}.

% Analysis

\begin{figure}[!htbp]
\centering
\includegraphics[width=8.4cm]{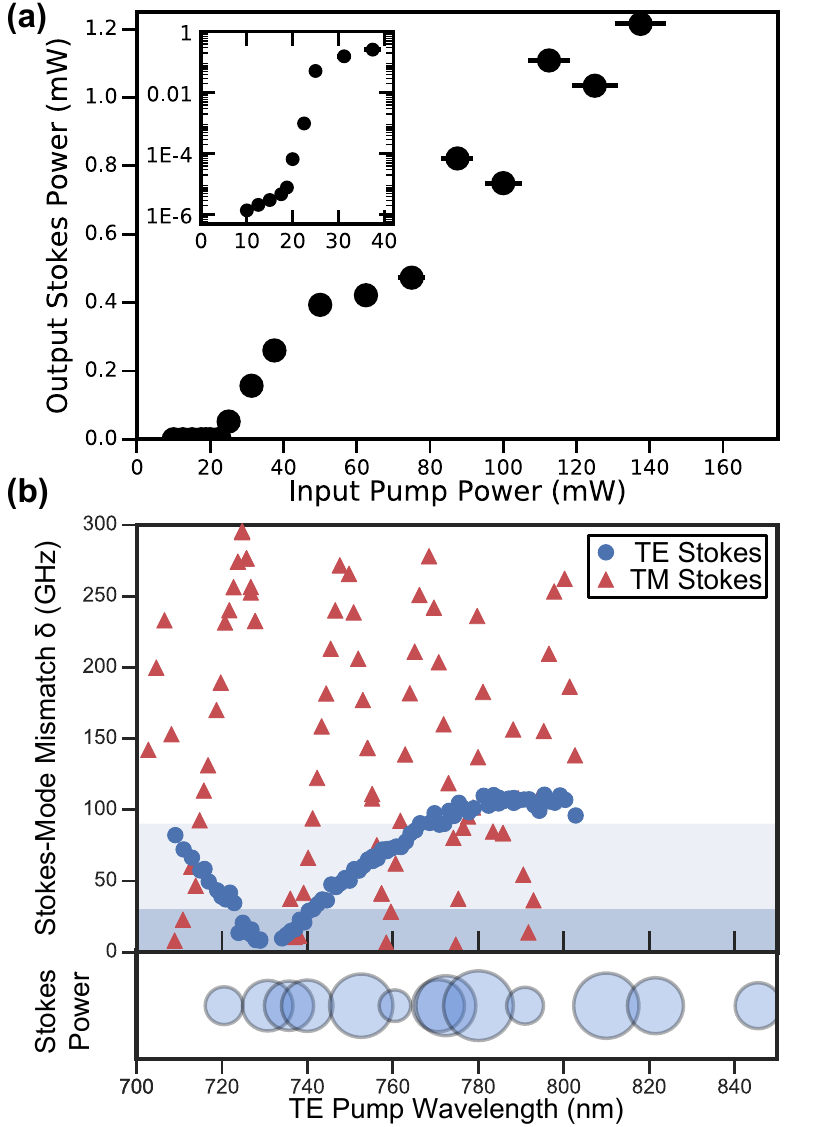}
\caption{Raman measurements of Device 1. (a) Threshold measurements for TE pump at the resonance from Fig. \ref{fig:resonances}, or $\SI{749.88}{\nm}$ and a TE-polarized Stokes at $\SI{834.34}{\nm}$. The inset shows a semi-log plot of the near-threshold data. (b) Stokes-mode mismatch plot (top) shows the mismatch between the Raman gain from a TE pump resonance and the nearest TE (blue, circles) and TM (red, triangles) resonance. The blue and light blue bands show the mode mismatch lying within one half-maximum and three half-maximums of the Raman gain peak, respectively. The ribbon (bottom) qualitatively shows (by circle size) recorded Stokes power from Raman lasing at the respective pump wavelengths. The TE-TE transition is well-matched across the measurement, and Stokes output is observed over $\SI{60}{\THz}$ of tuning (bottom).}
\label{fig:threshold}
\end{figure}

% Raman measurements - basic
Threshold measurements for TE pump and TE Stokes lasing (TE-TE process) in Device 1 are reported in  \ref{fig:threshold}(a). No higher-order Stokes or anti-Stokes processes were observed. With a Raman threshold of $\SI{20}{\mW}$, we calculate\cite{Spillane2002} an effective Raman gain value of $3.2 \ \si{\cm \per \GW}$. This is lower than previous values for diamond at these wavelengths ($>10 \ \si{\cm \per \GW}$\cite{Mildren}), and can be attributed to imperfect confinement ($74\%$ of the optical power is in the diamond core) and Raman gain dependence on the orientation of the crystal axes with respect to the propagation/polarization direction\cite{Mildren, Feigel2016}. The near-threshold data implies an external conversion efficiency of $1.7\%$, corresponding to an internal efficiency of $85\%$, close to the theoretical maximum\cite{Kippenberg2004a} of $90\%$. Furthermore, the Raman gain at a Stokes resonance may experience a Stokes-mode mismatch $\delta$, and possibly polarization conversion, as per Fig. \ref{fig:schematic}(d). Fig. \ref{fig:threshold}(b) shows the mismatch and recorded Stokes power at the various pump wavelengths. The locations of the resonances are recorded at low optical power, so thermal shifts are not taken into account. The bottom ribbon shows where Raman lasing was observed at high pump powers ($\SI{\sim200}{\mW}$ in the waveguide). Because the TE-TE transition was well-matched throughout the pump tuning range owing to suitable dispersion, Stokes lasing was observed with an output over a bandwidth of >150 nm, or $\SI{60}{\THz}$, corresponding to $17.5\%$ of the center frequency, from $800-\SI{950}{\nm}$.

\begin{figure*}[htbp]
\centering
\includegraphics[width=\columnwidth]{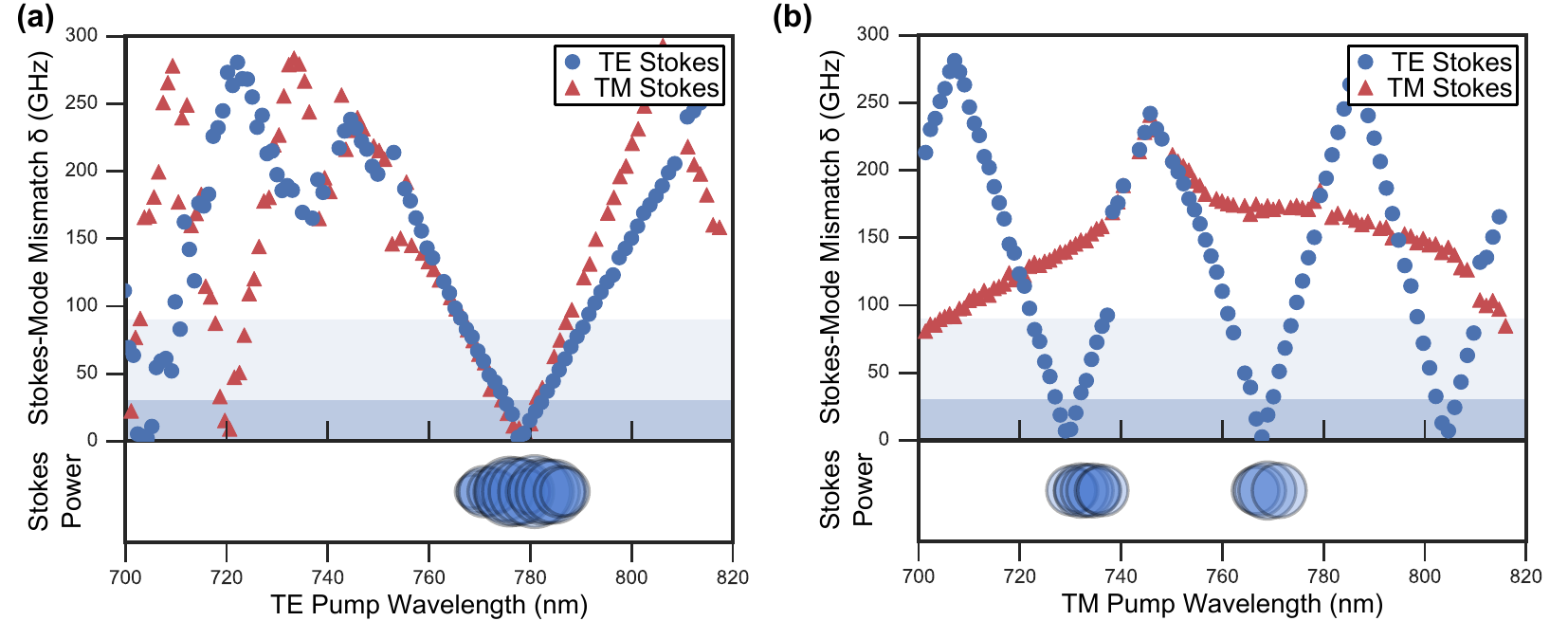}
\caption{Raman lasing and polarization conversion dependence on Stokes-mode mismatch for Device 2 with same nominal dimensions as Fig. \ref{fig:resonances}. (a) The mismatch between the Raman gain created by a TE pump resonance and the nearest TE (blue circles) and TM (red triangles) mode is plotted (top). Though both TE and TM resonances are well-matched, the output is likely TE, based on the higher quality factor of these resonances. (b) A similar plot as (a), but for TM pump resonance. Two regions of Stokes output are recorded at roughly $\SI{\sim735}{\nm}$ and $\SI{\sim770}{\nm}$. The TM pump is converted to a TE Stokes, and the neighborhood of modes for which Raman lasing is observed is narrower, as expected from the Stokes-mode mismatch plot.}
\label{fig:stokesmodematch}
\end{figure*}

Fig. \ref{fig:stokesmodematch} shows the Stokes-mode mismatch for both TE (a) and TM (b) pump wavelengths to TE Stokes (blue circles) and TM Stokes (red triangles) for the second device, which differs from the first due to fabrication variation. Stokes output is observed when the Stokes-mode mismatch is small for a given pump. Accordingly, lasing for a TE pump occurs over a relatively larger bandwidth, and lasing for a TM pump occurs in two distinct wavelength regions. The output is likely to be TE-polarized, due to the lower quality factors for TM resonances and forbidden TM-TM transition for [001]-oriented diamond\cite{Mildren, Feigel2016}.

\section{Conclusion}
% Summary results
In conclusion, we have demonstrated an integrated CW Raman laser pumped in the near-visible region. Owing to directly-written doped-glass waveguides, we were able to show lasing at high pump powers across 60 THz of tuning. The mismatch between the Raman gain peak and the nearest resonator mode was shown to be an important parameter, and polarization conversion from TM-TE was also observed. We studied the threshold behavior for TE-TE Raman lasing for a given Stokes output resonance, finding low threshold power $\SI{\sim20}{\mW}$.

% Closing remarks
These results strongly suggest the difficulty of suppressing Raman processes over a broad bandwidth in large-FSR resonators, as is required for generating visible Kerr frequency combs in diamond\cite{Okawachi2017a}. Nonetheless, the improvements in diamond quality factor and coupling-waveguide power handling are broadly applicable to integrated optics in the visible, including quantum, single-emitter based devices. The silicon-vacancy center, for instance, has an optical transition very close to the wavelengths studied here, and co-integration can be used for hybrid nonlinear-quantum optics devices. By translating these results to an angle-etched platform\cite{Burek2014, Latawiec2015b}, diamond's full potential can be realized.

\appendix
\section{Directly-written doped-glass waveguides}
\label{section:waveguidefab}
Due to the small size of the diamond substrate, waveguides which extend to the end of the sample cannot be natively written and defined. To couple light efficiently onto the diamond sample, we developed a doped spin-on glass process which allows for directly-writeable optical waveguides. This represents an evolution of previous polymer-based waveguides\cite{Hausmann2013}, as it has improved power-handling ability at visible wavelengths.

Alignment marks which were written with the diamond structures are first masked off with Kapton tape. A sol-gel solution is prepared by mixing FOx-16 and titanium butanate (Ti(OBu)$_4$) in a 4:1 ratio. The titanium butanate increases the refractive index of the final waveguides, ensuring the waveguides have a higher index than the substrate. Next, the diamond sample is placed on a spinner and brought up to 1000 RPM. Then, 2 drops of the sol-gel solution are dropped on the diamond sample as it begins to accelerate to 3000 RPM. The sample spins at 3000 RPM for 30 seconds. The sample is not baked, but undergoes an O$_2$ plasma treatment (300 mT, 75 W, 30 seconds) after the Kapton tape is removed to increase adhesion of conductive polymer to the masked area. To protect the deposited sol-gel during plasma treatment, a dummy piece of silicon is placed on top of the sample, exposing only the area where the tape was removed. The sample is then taken back to a spinner, where E-Spacer (ShowaDenko) is spun on at 4000 RPM. The waveguides are written at a dose of $6000 \si{\micro C / \cm^2}$. The sample develops for 10 seconds in TMAH, followed by a rinse in DI, then a rinse in methanol. To reduce absorption, the waveguides must be annealed. The anneal was limited to 460 C in O$_2$ for three hours to avoid burning the diamond. Ideally, a rapid thermal anneal to 1100 C in O$_2$ can completely oxidize the waveguides\cite{Holzwarth2007}, though the performance improvement over the low temperature anneal was negligible in this study.

\section{Funding}

Defense Advanced Research Projects Agency (DARPA) (W31P4Q-15-1-0013); National Science Foundation (NSF) (DGE1144152, ECS-0335765, DMR-1231319)

\section{Acknowledgement} This Letter was performed in part at the Center for Nanoscale Systems at Harvard, a member of the National Nanotechnology Infrastructure Network, supported by the NSF.

% Bibliography
\bibliographystyle{osajnl}
\bibliography{visraman}

\begin{thebibliography}{10}
\newcommand{\enquote}[1]{``#1''}

\bibitem{Mildren}
R.~Mildren and J.~Rabeau, \emph{{Optical Engineering of Diamond}} (Wiley,
  2013).

\bibitem{Mildren2009a}
R.~P. Mildren and A.~Sabella, \enquote{{Highly efficient diamond Raman laser.}}
  Opt. Lett. \textbf{34}, 2811--2813 (2009).

\bibitem{Williams2014}
R.~J. Williams, O.~Kitzler, A.~McKay, and R.~P. Mildren,
  \enquote{{Investigating diamond Raman lasers at the 100 W level using
  quasi-continuous-wave pumping},} Opt. Lett. \textbf{39}, 4152 (2014).

\bibitem{Granados2011}
E.~Granados, D.~J. Spence, and R.~P. Mildren, \enquote{{Deep ultraviolet
  diamond Raman laser.}} Opt. Express \textbf{19}, 10857--10863 (2011).

\bibitem{Mildren2008}
R.~P. Mildren, J.~E. Butler, and J.~R. Rabeau, \enquote{{CVD-diamond external
  cavity Raman laser at 573 nm.}} Opt. Express \textbf{16}, 18950--18955
  (2008).

\bibitem{Sabella2011}
A.~Sabella, J.~A. Piper, and R.~P. Mildren, \enquote{{Efficient conversion of a
  1064 {$\mu$}m Nd:YAG laser to the eye-safe region using a diamond Raman
  laser},} Opt. Express \textbf{19}, 23554 (2011).

\bibitem{Kitzler2017}
O.~Kitzler, J.~Lin, H.~M. Pask, R.~P. Mildren, S.~C. Webster, N.~Hempler,
  G.~P.~a. Malcolm, and D.~J. Spence, \enquote{{Single-longitudinal-mode ring
  diamond Raman laser},} Opt. Lett. \textbf{42}, 1229 (2017).

\bibitem{McKay2017}
A.~McKay, A.~Sabella, and R.~P. Mildren, \enquote{{Polarization conversion in
  cubic Raman crystals},} Sci. Rep. \textbf{7}, 41702 (2017).

\bibitem{Spillane2002}
S.~M. Spillane, T.~J. Kippenberg, and K.~J. Vahala,
  \enquote{{Ultralow-threshold Raman laser using a spherical dielectric
  microcavity.}} Nature \textbf{415}, 621--623 (2002).

\bibitem{Kippenberg2004a}
T.~J. Kippenberg, S.~M. Spillane, D.~K. Armani, and K.~J. Vahala,
  \enquote{{Ultralow-threshold microcavity Raman laser on a microelectronic
  chip.}} Opt. Lett. \textbf{29}, 1224--6 (2004).

\bibitem{Rong2005a}
H.~Rong, A.~Liu, R.~Jones, O.~Cohen, D.~Hak, R.~Nicolaescu, A.~Fang, and
  M.~Paniccia, \enquote{{An all-silicon Raman laser.}} Nature \textbf{433},
  292--294 (2005).

\bibitem{Rong2008}
H.~Rong, S.~Xu, O.~Cohen, O.~Raday, M.~Lee, V.~Sih, and M.~Paniccia,
  \enquote{{A cascaded silicon Raman laser},} Nat. Photonics \textbf{2},
  170--174 (2008).

\bibitem{Takahashi2013}
Y.~Takahashi, Y.~Inui, M.~Chihara, T.~Asano, R.~Terawaki, and S.~Noda,
  \enquote{{A micrometre-scale Raman silicon laser with a microwatt
  threshold.}} Nature \textbf{498}, 470--4 (2013).

\bibitem{Liu2017}
X.~Liu, C.~Sun, B.~Xiong, L.~Wang, J.~Wang, Y.~Han, Z.~Hao, H.~Li, Y.~Luo,
  J.~Yan, T.~Wei, Y.~Zhang, and J.~Wang, \enquote{{Integrated continuous-wave
  aluminum nitride Raman laser},} Optica \textbf{4}, 893 (2017).

\bibitem{Hausmann2014}
B.~J.~M. Hausmann, I.~Bulu, V.~Venkataraman, P.~Deotare, and M.~Lon{\v{c}}ar,
  \enquote{{Diamond nonlinear photonics},} Nat. Photonics \textbf{8}, 369--374
  (2014).

\bibitem{Latawiec2015}
P.~Latawiec, V.~Venkataraman, M.~J. Burek, B.~J.~M. Hausmann, I.~Bulu, and
  M.~Lon{\v{c}}ar, \enquote{{On-chip diamond Raman laser},} Optica \textbf{2},
  924--928 (2015).

\bibitem{Hausmann2012}
B.~J.~M. Hausmann, B.~Shields, Q.~Quan, P.~Maletinsky, M.~McCutcheon, J.~T.
  Choy, T.~M. Babinec, A.~Kubanek, A.~Yacoby, M.~D. Lukin, and M.~Lon{\v{c}}ar,
  \enquote{{Integrated diamond networks for quantum nanophotonics},} Nano Lett.
  \textbf{12}, 1578--1582 (2012).

\bibitem{Hiller2010a}
D.~Hiller, R.~Zierold, J.~Bachmann, M.~Alexe, Y.~Yang, J.~W. Gerlach,
  A.~Stesmans, M.~Jivanescu, U.~M{\"{u}}ller, J.~Vogt, H.~Hilmer,
  P.~L{\"{o}}per, M.~K{\"{u}}nle, F.~Munnik, K.~Nielsch, and M.~Zacharias,
  \enquote{{Low temperature silicon dioxide by thermal atomic layer deposition:
  Investigation of material properties},} J. Appl. Phys. \textbf{107}, 1--10
  (2010).

\bibitem{Osswald2006}
S.~Osswald, G.~Yushin, V.~Mochalin, S.~O. Kucheyev, and Y.~Gogotsi,
  \enquote{{Control of sp2/sp3 carbon ratio and surface chemistry of
  nanodiamond powders by selective oxidation in air},} J. Am. Chem. Soc.
  \textbf{128}, 11635--11642 (2006).

\bibitem{Feigel2016}
B.~Feigel, H.~Thienpont, and N.~Vermeulen, \enquote{{Design of infrared and
  ultraviolet Raman lasers based on grating-coupled integrated diamond ring
  resonators},} J. Opt. Soc. Am. B \textbf{33}, B5 (2016).

\bibitem{Okawachi2017a}
Y.~Okawachi, M.~Yu, V.~Venkataraman, P.~M. Latawiec, A.~G. Griffith, M.~Lipson,
  M.~Loncar, and A.~L. Gaeta, \enquote{{Competition between Raman and Kerr
  effects in microresonator comb generation},} Opt. Lett. \textbf{42},
  2786--2789 (2017).

\bibitem{Burek2014}
M.~J. Burek, Y.~Chu, M.~S.~Z. Liddy, P.~Patel, J.~Rochman, S.~Meesala, W.~Hong,
  Q.~Quan, M.~D. Lukin, and M.~Lon{\v{c}}ar, \enquote{{High quality-factor
  optical nanocavities in bulk single-crystal diamond},} Nat. Comm. \textbf{5},
  5718 (2014).

\bibitem{Latawiec2015b}
P.~Latawiec, M.~J. Burek, V.~Venkataraman, and M.~Lon{\v{c}}ar,
  \enquote{{Waveguide-loaded silica fibers for coupling to high-index
  micro-resonators},} Appl. Phys. Lett. \textbf{108}, 031103 (2016).

\bibitem{Hausmann2013}
B.~J.~M. Hausmann, I.~B. Bulu, P.~B. Deotare, M.~McCutcheon, V.~Venkataraman,
  M.~L. Markham, D.~J. Twitchen, and M.~Lon{\v{c}}ar, \enquote{{Integrated
  high-quality factor optical resonators in diamond.}} Nano Lett. \textbf{13},
  1898--902 (2013).

\bibitem{Holzwarth2007}
C.~W. Holzwarth, T.~Barwicz, and H.~I. Smith, \enquote{{Optimization of
  hydrogen silsesquioxane for photonic applications},} J. Vac. Sci. Technol. B
  \textbf{25}, 2658 (2007).

\end{thebibliography}

% Full bibliography added automatically for Optics Letters submissions; the following line will simply be ignored if submitting to other journals.
% Note that this extra page will not count against page length

\end{document}